# Experimental observation of Kerr-Raman solitons in a normal-dispersion FP resonator

TIEYING LI[1], KAN WU[1,*], XUJIA ZHANG[1], MINGLU CAI[1] AND JIANPING CHEN[1]

[1]*State Key Laboratory of Advanced Optical Communication Systems and Networks, School of Electronic Information and Electrical Engineering, Department of Electronic Engineering, Shanghai Jiao Tong University, Shanghai 200240, China*

**Corresponding author: kanwu@sjtu.edu.cn*

**Abstract:** Dissipative Kerr nonlinear cavities attract intense interest due to their rich dynamics and capability to generate broadband, low-noise mode-locked optical frequency combs for applications in optical communications, dual-comb spectroscopy, and photonic lidar, etc. Different from the Kerr effect, which is an electronic response, stimulated Raman scattering (SRS) is a delayed response to molecular vibrations in materials. In microcavities, when driven in an anomalous group velocity dispersion (GVD) regime, SRS typically leads to self-frequency shift of solitons and generation of breather solitons which have been verified both theoretically and experimentally. However, when driven in a normal GVD regime, recent theoretical work predicts that SRS can cause the locking of switching waves (SWs) and thus support bright moving localized structures (LSs), which we term as Kerr-Raman solitons (KRSs). Limited by the design of suitable experimental parameters, experimental observation of the KRSs is not achieved yet. Here, we provide numerical investigation, and to our knowledge, the first experimental observation of these SRS enabled KRSs in a fiber Fabry-Perot (FP) resonator with ultra-low normal GVD. Such Kerr-Raman solitons exhibit localized temporal features with strong oscillations at ~13 THz local frequency on the top of a flat-top pulse. The corresponding spectrum is a low-noise and broadband Kerr comb with typical platicon-like spectrum in the center and two Raman Stokes and anti-Stokes peaks located near ±13 THz away from the center. With such SRS enabled broadband Kerr comb, we have achieved a KRS spectrum with a repetition rate of ~3.68 GHz and a -40 dB spectral width of 260 nm. The corresponding comb tooth count is >9000, covering the S+C+L telecommunication bands. Moreover, the formation process of such KRSs is also revealed, and it is found that the GVD plays a key role in its generation. Our work will help to advance the study of the dynamics of optical frequency combs under the influence of SRS, as well as providing a broadband coherent mode-locked optical source for wide applications.

## I. INTRODUCTION

Dissipative Kerr solitons (DKSs) can be self-organized in high-Q Kerr nonlinear cavities based on a double balance of group velocity dispersion (GVD) and Kerr effects, and coherent pumping as well as intra-cavity dissipation [1]. In the frequency domain, the DKS corresponds to a Kerr optical frequency comb with equal frequency spacing. Due to its compactness, high repetition frequency, and good comb coherence, it has been widely used in recent years for applications such as optical communications [2], low noise microwave generation [3], ranging lidar [4, 5], dual-comb spectroscopy [6], etc. DKS was first discovered and applied in fiber rings as an optical buffer [7] and then emerged in various integrated platforms including $MgF_2$[8], $Si_3N_4$ [9-11], silica [12-15], $LiNbO_3$[16, 17], AlN [18], etc. These fiber rings, whispering gallery mode (WGM) cavities, or micro-rings usually require anomalous GVD to satisfy the spontaneous four-wave mixing phase matching condition. The DKSs are, to some extent, limited by thermal instability, low conversion efficiency, and low comb energy due to their presence in the red detuned regime of the bistable curve [19]. The material dispersion of most materials is usually normal GVD in the NIR, so additional waveguide dispersion designs are required to meet the anomalous GVD requirements. In the normal dispersion regime, self-organized structures called dark pulses [20, 21] exist. These dark pulses follow the upper branch of the bistable curve and have a high-duty cycle with high conversion efficiency, which can be seen arising from the interlocking of the bottom oscillations of the up-switching wave (SW) and the down-SW [22]. The bifurcation structure shows that the stable regime of dark pulses decreases with their different orders, which is known as the collapsed snake structure [23]. The platicons with flat-top pulse profiles [24-26], usually regarded as higher-order dark pulses, are accessible at the stable Maxwell point. Due to the absence of modulation instability at low detuning in the normal dispersion regime [27], the excitation of dark pulses need to be realized by employing avoided mode crossing [28-30], amplitude-modulated optical pumping [31-33], spectral filtering [34], self-injection locking [35], pulsed pumping [36], double-cavity coupled photonic molecules [37, 38], or photonic crystal microcavity [39] in the experiment.

In addition to the Kerr effect in a microcavity, the nonlinear scattering, stimulated Raman scattering (SRS), caused by the delayed response of the molecular vibrations also changes the dynamical properties and spatiotemporal stability of the field. When driven in an anomalous GVD regime, the effect of SRS mainly manifests itself in three ways. First, SRS causes the soliton self-frequency shift [40], i.e., the center of the spectral envelope is shifted away from the pump. Second, SRS can introduce additional Hopf bifurcations thereby changing the stability of the solitons and leading to breather solitons [41, 42]. Third, the cavity with SRS can support the generation of Stokes solitons [43]. In contrast, when driven in a normal GVD regime, SRS affects the spatio-temporal stability of the platicons, leading to additional branching in the time domain [44] or the generation of breather solitons [45]. Meanwhile, The SRS can cause the top-locking of SWs, resulting in the formation of local structures (LSs). Third order dispersion (TOD) can play a similar role, but there are clear differences: the role of TOD is expressed as the modulation of the spectrum, while SRS can introduce additional spectral components; the top oscillation of TOD-enabled LSs is locked to the up-SW, while that of SRS-enabled LSs is locked to the down-SW[46, 47]. The effect of TOD has been confirmed in $Si_3N_4$ microrings [48], FP resonators [49] and fiber rings [50]. However, to the best of our knowledge, the SRS enabled LSs are not yet reported and investigated experimentally, due to the lack of suitable experimental parameters designed to effectively enhance the interaction between SRS and Kerr effects in the normal GVD regime. To emphasize the importance of the interaction, we term this kind of LSs as Kerr-Raman solitons (KRSs).

In this work, we report for the first time the dynamical evolution of the fields under the influence of SRS driven in the normal GVD regime, based on a flexible fiber-based Fabry-Perot (FP) resonator platform. The effect of SRS on the stability of platicons is experimentally investigated, showing clear evidence for the formation of KRSs based on the top binding of SWs. It is found that the dispersion plays a key role in the generation of KRSs – the value of GVD should be in a proper region to simultaneously enable the interaction between the Raman gain and conventional Kerr platicon spectrum. The KRSs feature a temporal oscillation locking to the top of down-SW and a flat broadband spectrum benefited from the amplification of two platicon wings by Raman gain. We obtain a KRS comb with a -40 dB bandwidth of 260 nm covering S+C+L telecommunication bands, and a comb tooth spacing of 3.6787 GHz. The corresponding comb tooth number is greater than 9000, which is much larger than that reported in normal GVD microcavities. The KRSs also exhibit low-noise operation, good long-term stability and spectral tunability by desynchronization. Our work reveals a novel localized structure of mode-locked combs and is essential to facilitate the fundamental studies of SRS participated Kerr combs.

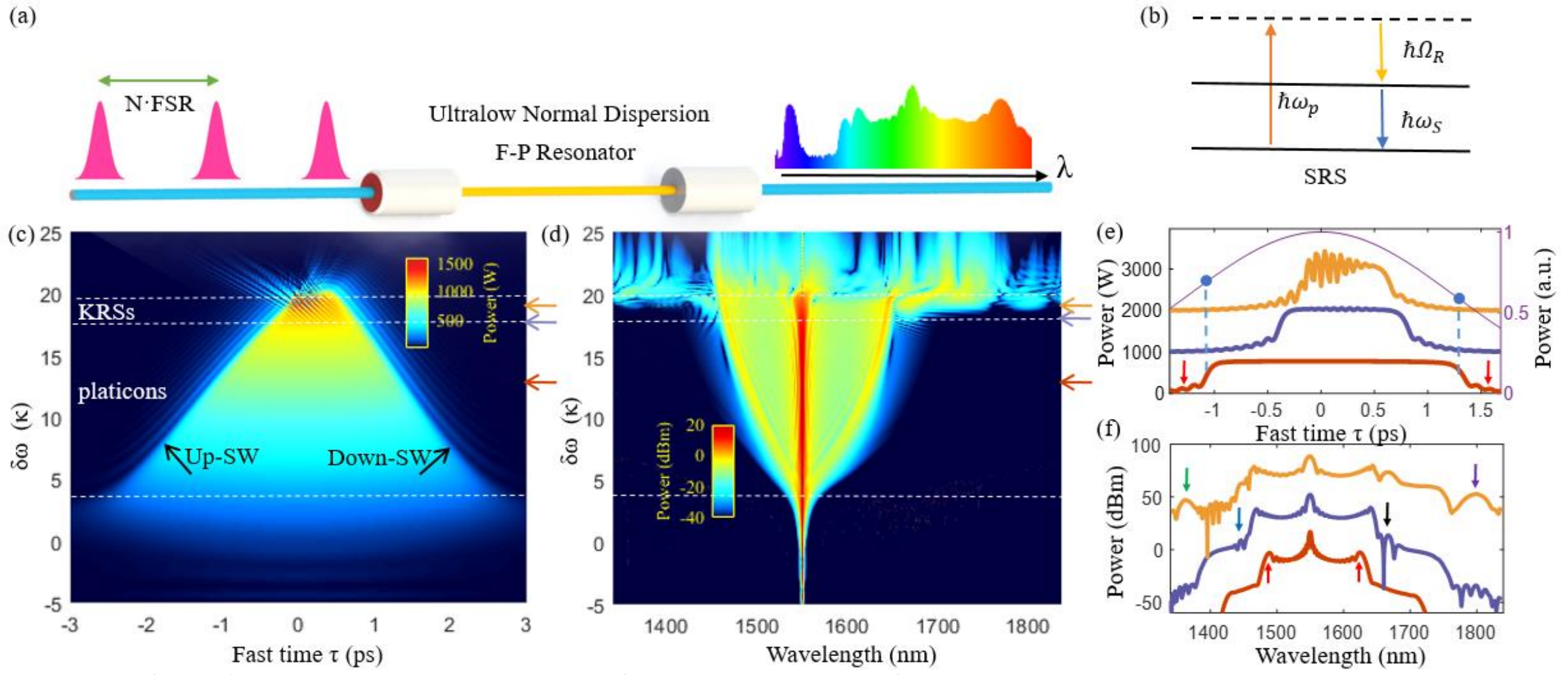

Fig. 1 Simulation of KRS formation. (a) Conceptual diagram for KRS generation in a fiber Fabry-Perot resonator with normal GVD and pulse pumping. (b) Energy level diagram for the SRS. (c) Temporal and (d) spectral evolution of the intra-cavity fields with respect to the detuning δω. The arrows (red, blue and yellow) correspond to δω = 13κ, 17.6κ and 18.9κ, respectively. The up-SW and down-SW are also denoted by black arrows in (c). (e) Corresponding temporal waveforms and (f) spectra marked by the arrows in (c) and (d). Purple line in (e): normalized profile of pump pulse. Arrows in (f): red – dispersive waves of platicons, blue and black – primary Stokes and anti-Stokes lights, purple and green - secondary Stokes and anti-Stokes lights

## II. THEORY

Figure 1(a) shows a schematic diagram for the study of KRSs in a normal dispersion FP resonator. The KRSs can be excited by pulsed pumping at a repetition rate which is an integer multiple of the cavity free spectral range (FSR) [51]. Two nonlinear processes take place in the cavity. One is the SRS (see Fig. 1(b)), where a pump photon is red-shifted to produce a Stokes photon with energy $\hbar\omega_S$ and release a phonon with energy $\hbar\Omega_R$. A wide range of SRS components can be generated due to the broad Raman gain spectrum of silica. The other is the four-wave mixing (FWM) dependent on the Kerr effect. The mutual interaction between FWM and SRS allows the generation of broadband frequency comb. To simulate the evolution of the internal optical field of the system under the combined the effect of SRS, Kerr effect and normal GVD, we start from the well-known Lugiato-Lefever equation (LLE) with a Raman scattering term and solve the equation using the Runge-Kutta method. The LLE can be expressed as follows[48]

$$\frac{\partial A(t,\tau)}{\partial t} = \mathcal{F}\left[i\left(\delta\omega+\mu\cdot 2\pi\delta f_{eo}+D_{\text{int}}(\mu)\right)\tilde{A}_{\mu}\right]-\frac{\kappa}{2}A + ig\left((1-f_R)|A|^2+f_R h_R(\tau)*|A|^2\right)A+\sqrt{\frac{\kappa_{ex}P_0}{\hbar\omega_0}}f_P(\tau) \quad (1)$$

where $A(t,\tau)$ describes the intra-cavity photon field, $t$ is the slow time, $\tau$ is the fast time, $\delta\omega$ is the detuning between the pump comb and the nearest resonance mode, $\delta f_{eo}$ is the mismatch between the pump repetition rate and the cavity FSR, $D_{\rm int} = \mu^2/2\cdot D_2+\mu^3/6\cdot D_3+\cdots$ is the linear phase operator describing the cavity dispersion, with the GVD coefficient $D_2$ and the TOD coefficient $D_3$, $\mu$ is the number of resonance mode, the loss rate $\kappa=\kappa_{ex}+\kappa_0$ includes the coupling rate $\kappa_{ex}$ and the intrinsic loss rate $\kappa_0$, $g$ denotes the nonlinear coupling coefficient, $f_R$ is the Raman fraction, $P_0$ is the peak power of pulse function $f_p(\tau)$. The SRS term is described by the corresponding time function $h_R(\tau)$, given by [52]

$$h_R(\tau) = \frac{\tau_1^2+\tau_2^2}{\tau_1\tau_2^2} e^{-\tau/\tau_2} \sin(\tau/\tau_1) \tag{2}$$

The values of these parameters are listed in the appendix table.

The intra-cavity field is excited by a Gaussian driving pulse $f_p(\tau)$ in the simulation, defined as $\exp(-\tau^2/\tau_p^2)$, with a pulse width $\tau_p$ = 2.5 ps, and a desynchronization frequency $\delta f_{eo} = 2\times$FSR- $f_{eo}$ = -15 kHz is introduced to tune the emission position of dispersive waves (DWs) to reach the Raman gain spectrum. The detuning $\delta\omega$ is scanned linearly from red to blue detuning, i.e., from -5$\kappa$ to 25$\kappa$, to drive the cavity through all possible states.

The evolution of the intra-cavity field in the time and frequency domain is shown in Fig. 1(c) and 1(d). Two typical optical states are identified during scanning, they are platicons and KRSs, and their regions are marked by the dashed lines. When scanning from 4$\kappa$ to 17.6$\kappa$, the intra-cavity field experiences a conventional platicon evolution. That is, the up-SW and the down-SW (marked by black arrows in Fig. 1(c)) can stabilize at the corresponding Maxwell points (specific pumping power for each detuning), e.g., the certain positions for $\delta\omega$ = 13$\kappa$ (marked by the blue dot in the time-domain envelope of the pump pulse, see Fig. 1(e) purple line), where the speed of the SWs can be regarded as almost zero relative to the pump pulse, thereby forming a stable platicon structure. If the SWs are pumped with a pumping power greater (or less) than the Maxwell point, the platicon temporal profile would expand (shrink) until the SWs stabilize at the Maxwell point. As the detuning $\delta\omega$ is gradually increased, the pumping power required at the Maxwell point for each detuning gradually increases, so the platicon temporal profile would become narrow and the corresponding spectrum would gradually broaden (see the platicons region of Fig. 1(c) and 1(d)). The platicon temporal profile at $\delta\omega$=13$\kappa$ is plotted in Fig. 1(e) (red line), with the trailing tails (marked by red arrows in Fig. 1(e)) at the bottom of the pulse corresponding to the DWs in the two wings (marked by red arrows) of the spectrum in Fig. 1(f) (red line).

As the detuning $\delta\omega$ is increased beyond 17.6$\kappa$, the intracavity field enters the KRS state. An oscillation first occurs at the top of the down-SW under the influence of SRS, as shown in Fig. 1(e) (blue line). The primary Stokes (black arrow) and anti-Stokes (blue arrow) lights begin to appear and grow in the spectrum as shown in Fig. 1(f) (blue line). The strong peak of anti-Stokes light indicates the contribution of FWM between pump and Stokes light [44]. Then, by slightly increasing the detuning, the DW peaks become close enough to the Raman gain spectrum, i.e., ~13 THz from the pump wavelength, and act as a seed to the Raman amplification. As a result, the spectrum experiences a significant broadening due to the mutual interaction of SRS and cascaded FWM effect, as shown in Fig. 1(f) (yellow line, $\delta\omega$ = 18.9$\kappa$) with two secondary Stokes and anti-Stokes peaks denoted. In the time domain, a strong oscillation on the top of the flat-top pulse is observed, as shown in Fig. 1(e) (yellow line). This oscillation is locked to the down-SW (see Supplementary Section 3). The up-SW is bounding to the down-SW through the oscillation forming the KRS structure.

It has been confirmed that such a KRS state cannot be obtained without SRS (see Supplementary Section 1). As the detuning further increases to 19.5$\kappa$, the KRSs lose the stability and enter a chaotic state, then disappear.

## III. EXPERIMENTAL SETUP

The experimental demonstration of KRS state is performed in a fiber FP resonator with a cavity length of ~5 cm, as shown in Fig. 2(a). The corresponding FSR is measured to be ~1.8393 GHz by using a dual-cavity method [49] (see Supplementary Section 4). The cavity is made of a highly nonlinear fiber (Thorlabs HN1550). Two facets of the fiber are fixed in ceramic ferrules and then polished and coated with a highly reflective film of 25 pairs of $SiO_2$ and $Ta_2O_5$ (see Fig. 2(b)). The films can form a Bragg mirror with reflectivity up to 99.6%, and the entire FP resonator is fixed in a homemade copper fixture to control the temperature. Due to the low film absorption and fiber transmission loss (0.9 dB/km), the resonator linewidth is measured to be 19 MHz (see Fig. 2(c)), corresponding to a loaded Q factor of $1\times10^7$.

The experimental setup is shown in Fig. 2(d). An electro-optical comb (EO comb) source is used to pump the FP resonator at a repetition rate of 3.6787 GHz (~2×FSR). The light from the EO comb is first compressed by a dispersion compensation fiber (DCF) to 3.1 ps, measured by a 500 GHz optical sampling oscilloscope (OSO, Alnair Labs EYE-2000C) as shown in Fig. 2(f). And the EO comb spectrum is shown in Fig. 3(g). Then a second nonlinear compression stage is adopted. The light is amplified to 28 dBm by an erbium-doped fiber amplifier (EDFA) and fed to 100 m single mode fiber (SMF) for nonlinear spectral broadening and pulse compression. The finally obtained pulse duration is 2.58 ps as shown in Fig. 3(h) and the corresponding spectral width is 7.4 nm as shown in Fig. 3(i). The repetition rate of the pulse $f_{eo}$ is set to 3.6787 GHz (~2×FSR) with a desynchronization frequency $\delta f_{eo}$ = 2×FSR - $f_{eo}$ = -13 kHz. The rate of the generated KRS comb will follow the pump pulse [51]. The average optical power coupled to the FP microcavity is estimated to 25 dBm due to fiber optic connection loss.

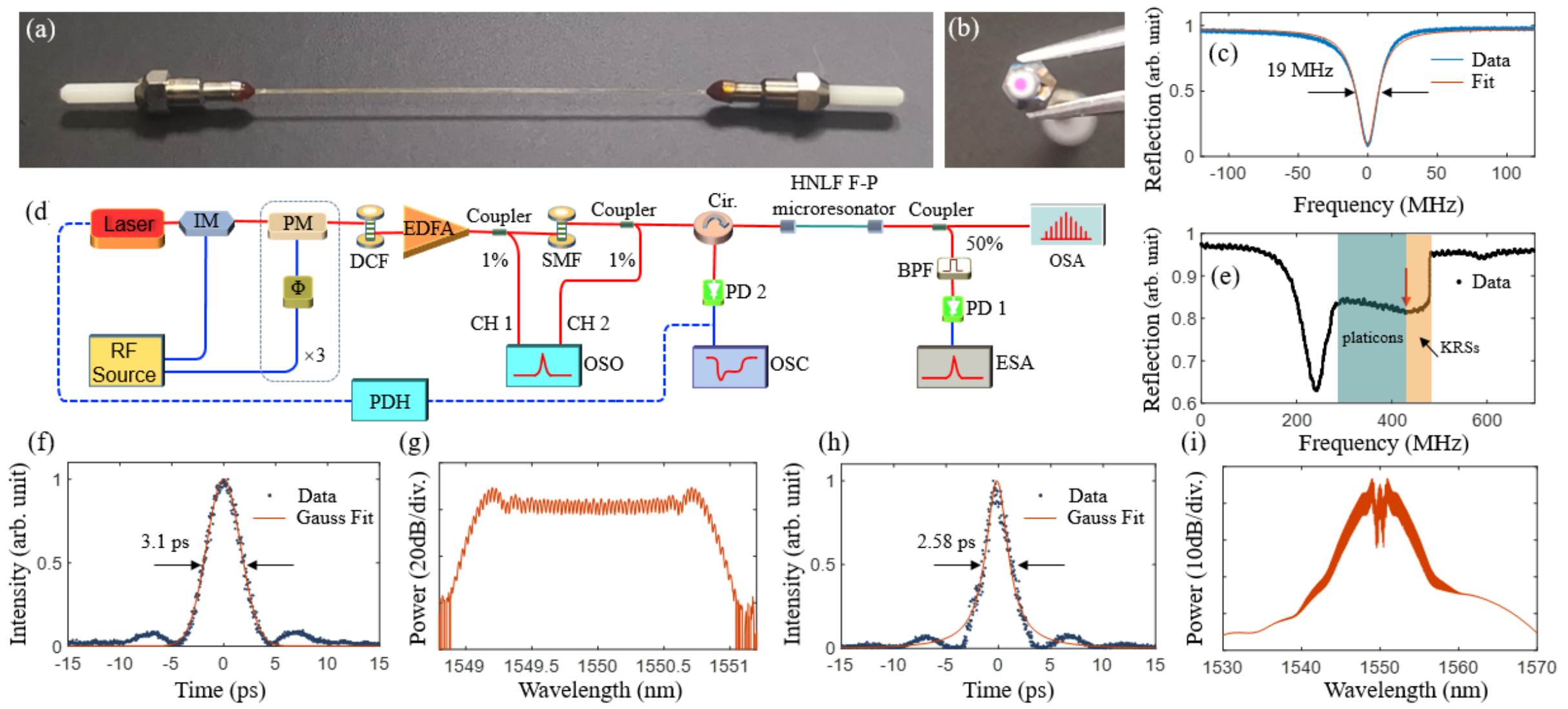

Fig.2 Experiment of KRS formation. (a) Photograph of the FP resonator. (b) Image of the reflective film at a facet of ceramic ferrule. (c) Reflection of the cold cavity near 1550 nm. (d) Experimental setup. IM: intensity modulator; PM: phase modulators; Φ: phase shifter; DCF: dispersion compensating fiber; SMF: single mode fiber; EDFA: erbium-doped fiber amplifier; OSO: optical sampling oscilloscope; PD: photodetector; BPF: band pass filter; OSC: oscilloscope; ESA: electronic spectrum analyzer; OSA: optical spectrum analyzer; PDH: Pound-Drever-Hall. (e) Resonance reflection at a pump tuning speed of 58 GHz/s. Green region: platicon state; yellow region: KRS state; red arrow: demarcation point. (f) Temporal waveform and (g) spectrum of the initial pump pulse. (h) Temporal waveform and (i) spectrum of the nonlinearly compressed pulse.

The dispersion parameters of the fiber are crucial for observing the KRSs. Firstly, the GVD and TOD should be small enough so that DWs can overlap with the Raman gain spectrum and stimulate the interaction of FWM and SRS (see Supplementary Section 2). Secondly, it is necessary to ensure that the normal GVD requirement is met both in the pump region and the entire Raman gain region for the generation of KRSs. Only conventional soliton is experimentally observed in a recent work[36], because the anomalous GVD in the Raman gain region could hinder the formation of KRSs.

A Mach-Zehnder interferometer is used to characterize the dispersion profile $D_{int} = D_2 \cdot \mu^2 + D_3 \cdot \mu^3$ of the FP resonator by sweeping the position of the resonance mode of the FP resonator. The details can be found in the Supplementary Section 5. The $D_2$ and $D_3$ are estimated as $2\pi \cdot (-28.2)$ Hz and $2\pi \cdot 5.2$ mHz, respectively.

## IV. EXPERIMENTAL OBSERVATION OF KRS

To explore the comb evolution of the cavity, the laser frequency is first swept across a resonance mode from the blue–detuned side to the red-detuned side at a high speed of 58 GHz/s. The light power reflected from the FP resonator detected by PD2 is plotted in Fig. 2(e). A step feature can be clearly observed. From the previous theoretical analysis, the intracavity field first enters a platicon state whose intracavity power (reflected power) increases (decreases) with the increase of detuning. Then the state transits to KRS state whose intracavity power (reflected power) decreases (increases) with the further increase of detuning. The demarcation point is indicated by a red arrow in Fig. 2(e).

As the platicon and KRS states are generated in the effectively blue-tuned region in a normal GVD microcavity, pulsed pumping scheme (with a weaker thermal effect) allows us to scan the pump frequency at a very low speed [15]. To effectively record the spectral evolution of the platicon and KRS states, we then adopt a slow detuning speed of 1.7 MHz/s by programmatically varying the laser's internal PZT voltage and simultaneously use an optical spectrum analyzer (OSA, AQ6370D) to record the spectra and a photodetector (PD2) to record the reflected light power.

The recorded spectrum is shown in Fig. 3(a), where the horizontal axis represents the wavelength and the vertical axis represents the frequency offset relative to the initial laser frequency calculated from the PZT voltage change. Note that the offset is the absolute offset of the laser frequency, not the offset relative to the resonance mode due to the red shift of the resonance mode with tuning. The initial tuning position is defined as the start position (0 MHz) and forward tuning is performed by decreasing the laser frequency at a tuning rate of 1.7 MHz/s. The output spectra (Fig. 3(a)) are gradually broadened during forward tuning, while the reflected light power (Fig. 3(b)) is gradually reduced, i.e., more pump power is coupled into the cavity. When the detuning exceeds -650 MHz, a step feature occurs with a length of 185 MHz, as shown in Fig. 3(b). The intracavity state enters the KRS state due to the appearance of Stokes light. According to the analysis in Fig. 1(c) the intracavity state first enters the platicon state but it stops at a KRS state. This is because the response speed of the thermal effect is much faster than the tuning speed, when we visit the spectral state on the step, the resonance mode will move towards the blue side due to the sudden power reduction in the cavity,

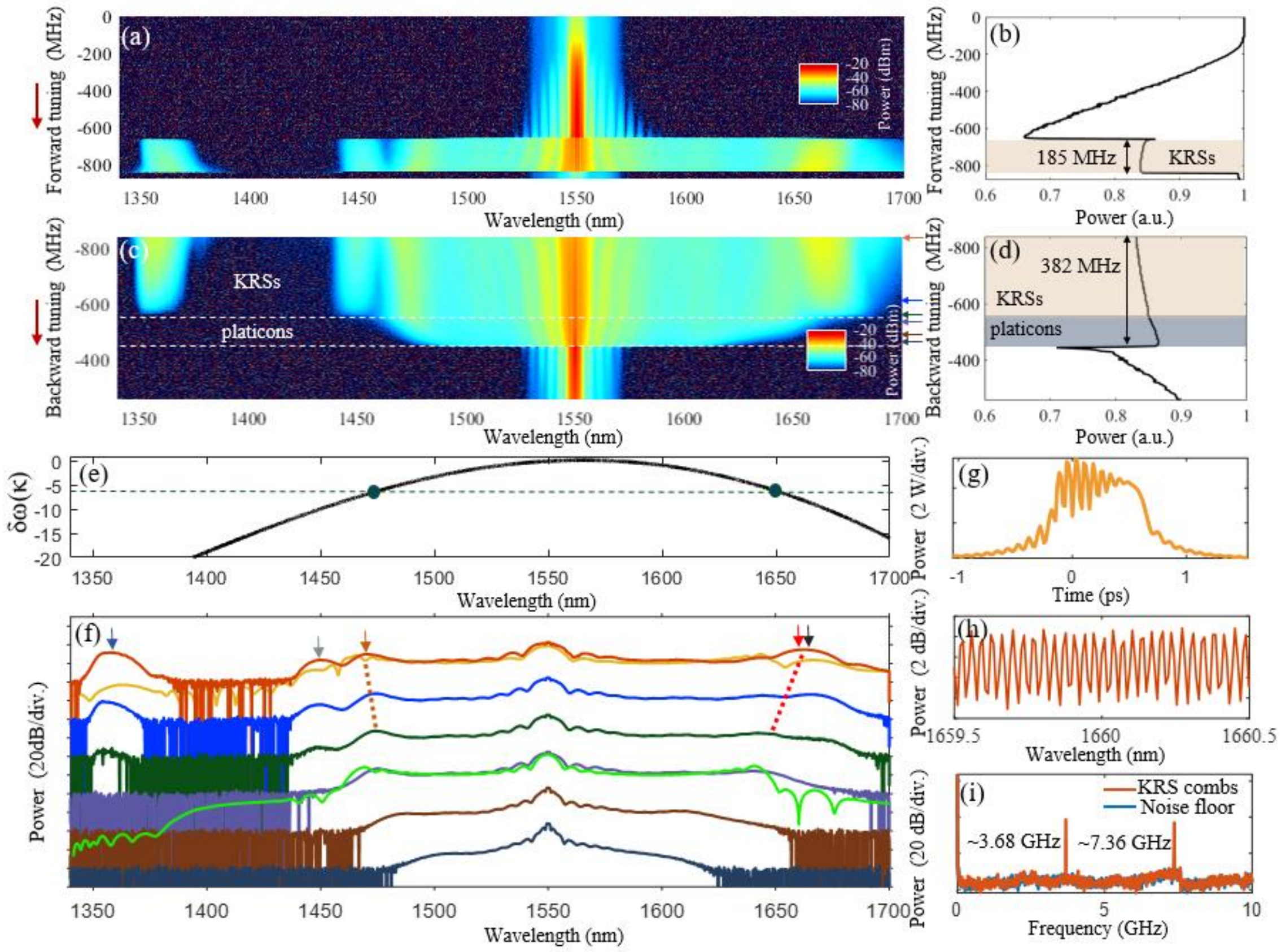

Fig. 3 The spectra evolution for the formation of the KRSs: (a) Experimental evolution of the spectrum and (b) reflected power vs. forward tuning. (c) Experimental evolution of spectrum and (d) reflected power vs. backward tuning. (e) Curve of $D_{int}+2\pi\cdot\delta f_{eo}$ profile to calculate the position of DWs. (f) Comb spectra corresponding to the arrows in (c) (red, blue, dark green, indigo, brown, and deep blue lines, correspond to -836.0, -608.0, -564.2, -546.6, -476.4, -447.2 MHz from top to bottom), and the simulation spectra (yellow and light green). (g) Simulated temporal profile of a KRS state corresponding to red line in (f). (h) Enlarged individual comb lines near 1660 nm. (i) Beatnote of the KRS comb detected by a 10-GHz PD.

resulting in only the back edge of the step, i.e., KRS state, being addressed.

To explore the field state of the entire step, a similar backward tuning method [53] is used. First, the laser frequency is tuned to ensure that the cavity outputs the widest KRS comb state (Fig. 3(c)), where the detuning is defined as the same value with the same state in Fig.3 (a). Then, backward tuning is performed, i.e., the laser frequency is increased. Benefiting from the weaker thermal influence during backward tuning [53], the step width increases to 382 MHz as shown in in Fig. 3(d). The regions of KRS and platicon states are marked in Fig. 3(c). Three spectra (red, blue and dark green lines correspond to the arrows in Fig. 3(c)) of KRS state with different detuning (-836.0 MHz, -608.0 MHz, -564.2 MHz) are plotted in the top three lines of Fig. 3(f). These spectra cover the range from 1350 nm to 1700 nm with a comb spacing of 3.6787 GHz. The enlarged comb spectrum near 1660 nm is shown in Fig. 3(h).

In the widest KRS comb spectrum (red line in Fig. 3(f)), five peaks (marked by different color arrows) are identified. From left to right, these peaks are secondary anti-Stokes light, primary anti-Stokes light, dispersive wave 1 (DW1), DW2, and primary Stokes light. The high power of secondary anti-Stokes light indicates the power transfer by FWM from pump and primary anti-Stokes light. The detection of secondary Stokes light is limited by the detection range of the OSA (600 nm-1700 nm). The shift of DW1 and DW2 with detuning is marked by dashed lines. Note that the positions of DW2 and the primary Stokes light overlap, which would be separated with backward detuning. The position of DW1 and DW2 can be derived from the formula $D_{int}+\mu\cdot 2\pi\cdot\delta f_{eo}+\delta\omega = 0$. The predicted position of DWs is marked by green dots in Fig. 3(e) when $\delta\omega$ = -6.2$\kappa$, which is consistent with the measured position of DWs in dark green line in Fig. 3(f). The desynchronization frequency $\delta f_{eo}$ can provide additional degrees of freedom to tune the spectra, which will be demonstrated in subsequent experiments. The simulation results (yellow and light green lines) show good agreement with the experimental spectra (red and indigo lines) at desynchronization frequency $\delta f_{eo}$ = -15 kHz.

The simulated temporal profile of the widest KRS state (red line in Fig. 3(f)) is shown in Fig. 3(g). The SRS induced strong oscillation can be clearly observed on the top of a flat-top pulse. It should be emphasized again that this oscillation is locked to the down-SW, i.e., trailing edge of the pulse (see Supplementary Section 3). The oscillation period is 74.6 fs corresponding to the Raman gain peak frequency ~13 THz. The small ringing tails before and after the pulse are caused by DW2 and DW1. Note that the KRS state is self-stable and can be trapped at a position on the trailing edge of the pump pulse similar to the zero-dispersion soliton [48], which distinguishes it from the platicon state in which the SWs depend on the Maxwell point.

The beat note of this widest KRS comb (red line) is measured directly with a 10-GHz PD, as shown in Fig. 3(i). It can be observed that there only exist repetition frequency of 3.6787 GHz and its harmonics, indicating that all the comb lines in the KRS spectrum belong to the same mode family, which is different from the Stokes soliton [43]. This result supports our analysis that the DWs of the platicon act as seed and are amplified by the gain of SRS, and finally the broadband KRS comb is generated by the mutual interaction of FWM and SRS.

The spectra of the platicon state at different detuning values (-546.6 MHz, -476.4 MHz, -447.2 MHz) are plotted in the bottom three lines (indigo, brown, and deep blue) in Fig. 3(f). With the backward detuning, the spectral width (span between two DWs) of platicon combs varies from 170 nm to 100 nm experimentally.

## V. CHARACTERISTICS OF PLATICON AND KRS COMBS

To confirm the low-noise operation of platicon and KRS states, we measure their noise performance. The combs are first amplified by an EDFA, and then a portion of the combs near 1560 nm is filtered out by a band-pass filter (7 nm bandwidth). The beatnotes of the filtered combs are measured by a high speed PD (FINISAR 40 GHz). The beatnotes of both states have a signal-to-noise ratio (SNR) of 45 dB, as shown in Fig.4(a). The platicon signal (blue line) has a slightly lower amplitude due to the weaker power of the comb compared to the KRS (red line). The phase noise spectra of these two states are also measured, as shown in Fig. 4(b). It can be found that two states have nearly identical noise spectra with values of -100 dBc·Hz$^{-1}$ at 1 kHz and -110 dBc·Hz$^{-1}$ at 10 kHz. This noise level is better than the work using the similar fiber FP resonators with repetition rate of 10 GHz, e.g. near-zero-dispersion soliton [49] (-85 dBc·Hz$^{-1}$ at 1 kHz, -104 dBc·Hz$^{-1}$ at 10 kHz,). The low-frequency noise of different KRS and platicon states are also measured by an APD (bandwidth 200 MHz) and electrical spectrum analyzer, as shown in Fig. 4(c). The colored lines correspond to the comb states with same color in Fig. 3(f). It can be seen that the noise slightly increases in the KRS states, which is attributed to the noise introduced by the Raman amplification [54].

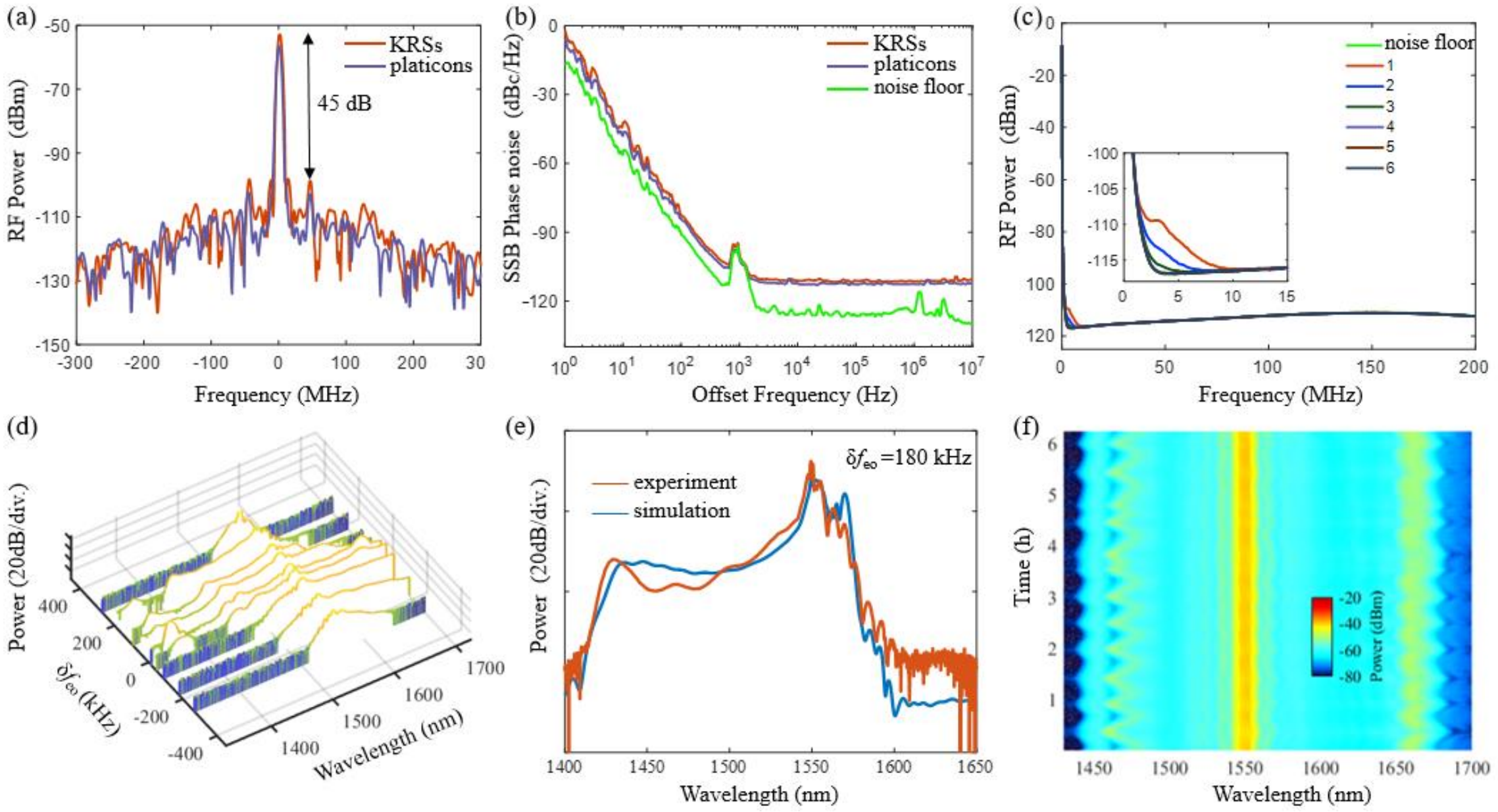

Fig. 4 Characteristics of the optical frequency combs: (a) Beatnote signals of platicons (blue) and KRSs (red) near 1560 nm. (b) Phase noise spectra of platicons (blue), KRSs (red), and pump light (green). (c) Low-frequency noise of different spectra (No. 1-6 lines correspond to the same color spectra in Fig. 3(f)). Inset: a zoomed view of the spectra below 15 MHz. (d) Spectral tunability vs. different $\delta f_{eo}$. (e) Experimental (red) and simulated (blue) spectra at $\delta f_{eo}$ = 180 kHz (f) Measured long-term spectral stability with PDH locking

The platicon spectrum be tuned experimentally by the desynchronization frequency $\delta f_{eo}$. In the experiments, the pump repetition frequency is fixed and the cavity FSR is finely tuned by changing the cavity temperature. The changing spectra at different $\delta f_{eo}$ are plotted in Fig. 4(d). The frequency combs exist in the range of $\delta f_{eo}$ from -270 to 270 kHz. When $\delta f_{eo}$ is positive, the spectra are shifted towards the short wavelength (e.g., span: 1400 nm ~ 1600 nm at $\delta f_{eo}$ = 180 kHz). When $\delta f_{eo}$ is negative, the spectra are shifted towards the long wavelength (e.g., span: 1500 nm ~ 1700 nm at $\delta f_{eo}$ = -180 kHz), and the maximum wavelength tuning range is up to 100 nm. The simulated spectrum shows a good agreement with the experimental result at $\delta f_{eo}$ = 180 kHz, as shown in Fig. 4(e).

The long-term stability of the KRS combs is crucial for practical application scenarios. Switching between platicon and KRS states causes a dip in the soliton step of reflected light power (marked by the red arrow in Fig. 2(e)), which allows the Pound-Drever-Hall (PDH) technique to be used to stabilize the frequency at a fixed detuning relative to the resonant mode. A digital laser locking module (TOPTICA DigiLock110) is used to modulate the laser current to generate sidebands and the reflected light is sent to the module to generate error signal to lock the detuning. When unlocked, the KRS comb typically loses its state after a few minutes due to the environmental perturbation even though it is thermally stable. Once locked, the KRS comb can be stable for more than 6 hours, as shown in Fig. 4(f).

## VI. CONCLUSION

In summary, we report the first direct experimental observation of a Kerr-Raman soliton state based on a fiber FP resonator platform made of low normal GVD fiber. This KRS state has a featured oscillation locked to the down-SW in the time domain and a broadband SRS enabled comb spectrum in the frequency domain. Experimentally, we have obtained a KRS comb with a repetition rate of ~3.68 GHz, a spectral width of 260 nm, and a comb line number > 9000. Compared with the conventional platicons, the KRS combs show excellent performance in terms of bandwidth and comb line number. (see Supplementary Section 6) The rich dynamics of field evolution under the mutual interaction of FWM, SRS and dispersion is revealed both in simulation and in experiment. The low-noise property, spectral tunability and long-term operation are also demonstrated. Our work facilitates the fundamental studies of SRS participated Kerr comb evolution as well as paving a way for a novel stable and tunable optical frequency comb source for wide applications.

### Appendix A: parameter values

| Parameter | Experimentally measured values or given values by other work | Values used in simulation |
|---|---|---|
| $D_2/2\pi$ ($\beta_2$) | -28.2±10 Hz (1.0 $ps^2$/km) | -36 Hz (1.3 $ps^2$/km) |
| $D_3/2\pi$ ($\beta_3$) | 5.2±2 mHz 0.0149 $ps^3$/km | 4 mHz 0.0115 $ps^3$/km |
| $\gamma$ | 10.8 $W^{-1}$•$km^{-1}$ | 10.8 $W^{-1}$•$km^{-1}$ |
| $\kappa/2\pi$ | 19±2 MHz | 20 MHz |
| $P_0$ | 21.5±2 W | 20 W |
| $\tau_p$ | 2.58±0.1 ps | 2.5 ps |
| $\delta f_{eo}$ | -13±2 kHz | -15 kHz |
| $P_0$ | 21.4±2W | 20 W |
| $\tau_1$ | 12.2 fs | 12.2 fs |
| $\tau_2$ | 32 fs | 32 fs |
| $F_R$ | 0.18 | 0.2 |

### Funding.

This work is supported by National Nature Science Foundation of China (NSFC) (No. 61922056).

### Disclosures.

The authors declare that they have no conflict of interest.

See Supplement material for supporting content.

# Experimental observation of Kerr-Raman solitons in a normal-dispersion FP resonator: Supplemental Materials

**TIEYING LI[1], KAN WU[1,*], XUJIA ZHANG[1], MINGLU CAI[1] AND JIANPING CHEN[1]**

[1]*State Key Laboratory of Advanced Optical Communication Systems and Networks, School of Electronic Information and Electrical Engineering, Department of Electronic Engineering, Shanghai Jiao Tong University, Shanghai 200240, China*
**Corresponding author: kanwu@sjtu.edu.cn*

## S1 Simulation of LLE without SRS

To further reveal the effect of SRS, the temporal (Fig. S1(a)) and spectral (Fig. S1(b)) evolutions of the field are investigated in the simulation of LLE without SRS term. It is obvious that only the platicon state exists during the field evolution due to the absence of SRS. TOD can play a similar role like SRS, when the TOD is large enough [1]. However, $d_3$ ($d_3 = 2/\kappa \cdot D_3/6 \cdot |\kappa/D_2|^{1.5}$ is the contribution of TOD relative to GVD) in the manuscript is calculated as 0.062, which is much smaller than 2.11 of the study [1], therefore the influence of TOD can be ignored and the generation of KRSs is attributed to the SRS rather than the TOD in our work.

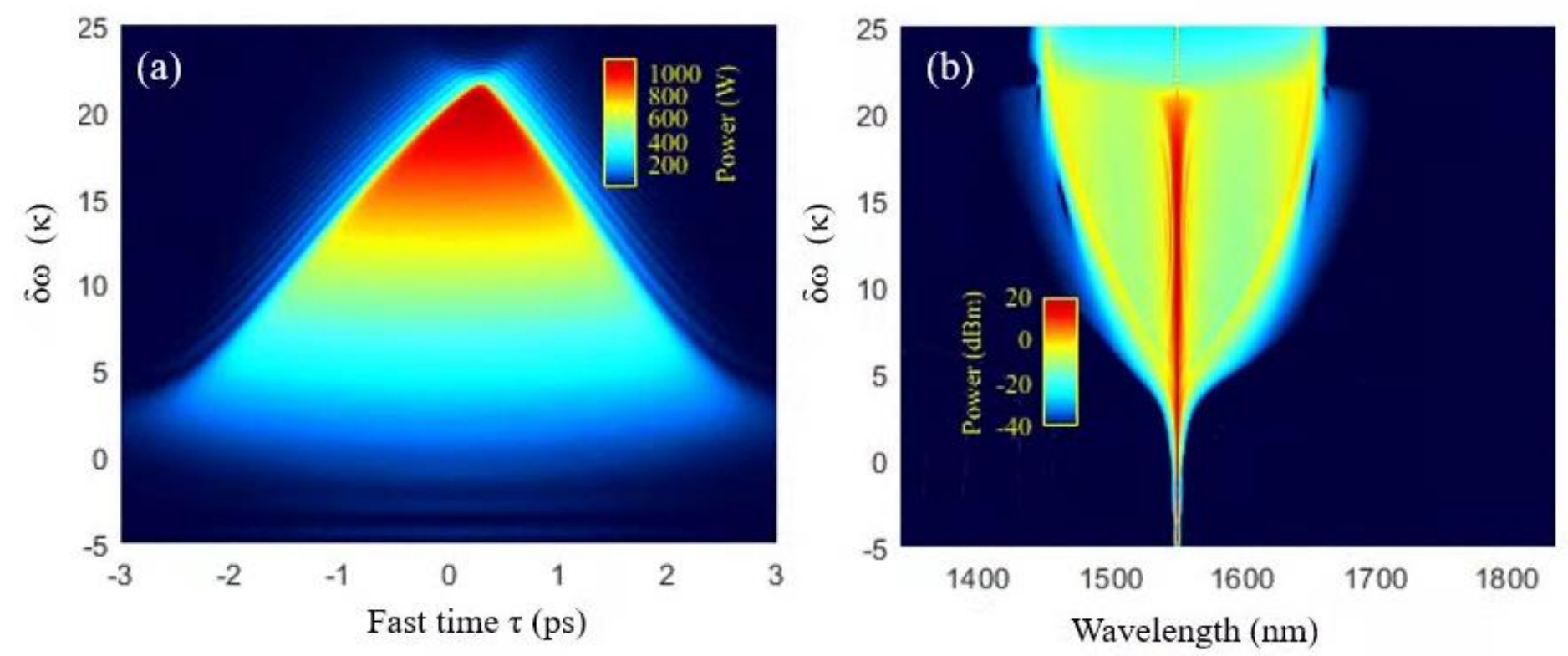

Fig. S1 LLE simulation without SRS: (a) Temporal and (b) spectral evolution without the influence of SRS.

## S2 Influence of GVD and TOD on of the KRS existence region

The existence region of KRSs is mainly influenced by the dispersion including the GVD (i.e., $\beta_2$ or $D_2$) and the TOD (i.e., $\beta_3$ or $D_3$). The normalized parameter ranges are determined through simulation to guide the dispersion design of the microcavity used to observe KRSs in the experiment. We first define the normalized parameters of fast time $\tau'$, field amplitude $\psi$, TOD $d_3$, detuning $\zeta_0$ and pump intensity $S$ as follows [1-3]:

$$\tau' = \tau \times D_1 \times \sqrt{\left|\frac{\kappa}{D_2}\right|};\quad \psi = A \times \sqrt{\frac{2g_0}{k}};\ d_3 = \frac{2}{k} \times \frac{D_3}{6} \times \left(\frac{\kappa}{D_2}\right)^{\frac{3}{2}};\ \zeta_0 = \frac{2\delta w}{\kappa};\ S = \frac{8\kappa_{ex} g_0}{\kappa^3 \hbar w_0} p$$

where $\tau$ is the fast time; $D_1 = 2\pi$ FSR; $\kappa$ is the loss rate; $D_2$ is the GVD; $A$ is the field amplitude; $g_0$ is nonlinear coupling term; $D_3$ is the TOD; $\delta\omega$ is the detuning between the driving carrier frequency $\omega_0$ and the nearby resonance; $\kappa_{ex}$ is the coupling rate; $p$ is the peak power of the pump pulse.

The GVD range of the KRS existence region is found to be between 1.8 ps$^2$/km and 1 ps$^2$/km when the normalized detuning $\zeta_0$ are fixed to 37.8, and the TOD are fixed to 0.0115 ps$^3$/km. When the GVD is higher than 1.8 ps$^2$/km, two wings of the platicon spectrum (i.e., the DWs) are not wide enough to enhance the interaction with SRS effect. As shown in Fig. S2(a-c), we can find that as the GVD decreases, the KRSs can be observed under a wider range of normalized pump power $S$, i.e., a wider KRS existence region. When the GVD is less than 1 ps$^2$/km, two wings of platicon spectrum （i.e., DWs）generated by FWM progress change their role from cooperation to competition with SRS [4]. The FWM progress is stronger in competition with SRS, which will hinder their interaction in the Raman gain region. The spectra and temporal waveforms of the KRSs at critical values of $S$ are shown in Fig. S2(1-6), corresponding to the lower and higher boundaries in Fig. S2(a-c). In Fig. S2(a-c), it is found that the pulse width of the KRSs increases with the increased pump power, which is similar to the evolution of a conventional platicon pulse [5]. The temporal and spectral profiles for each point in Fig.S2 can be found in the supplementary GIF image.

We also investigated the effect of TOD on the presence of KRSs at three different GVD values (Fig. S3(a-c)). To ensure that the dispersive waves (i.e., the wings of the platicon spectrum) are emitted in the Raman gain region, we introduce additional $\delta f_{eo}$ and made $\delta f_{eo}$ = -15 kHz when normalized TOD d3 >= 0 and $\delta f_{eo}$ = 15 kHz when d3 < 0. As shown in Fig. S3(a), the |d3| is limited to 0.17, when GVD is 1.0 ps$^2$/km (Fig. S3(a)). |d3| is reduced to 0.15, when the GVD is 1.3 ps$^2$/km (Fig. S3(b)). However, when the GVD is 1.8 ps$^2$/km, only two points can support the KRS solutions (Fig. S3(c)). This is because $d_3$ causes a change in the $D_{int}$ curve, which affects the emission position of the DWs that act as Raman gain seeds. When the GVD is smaller, the power of DWs in the Raman gain spectrum is greater, so the effect of changing $d_3$ is weaker than in the case of a larger GVD, where the power of DWs is extremely small.

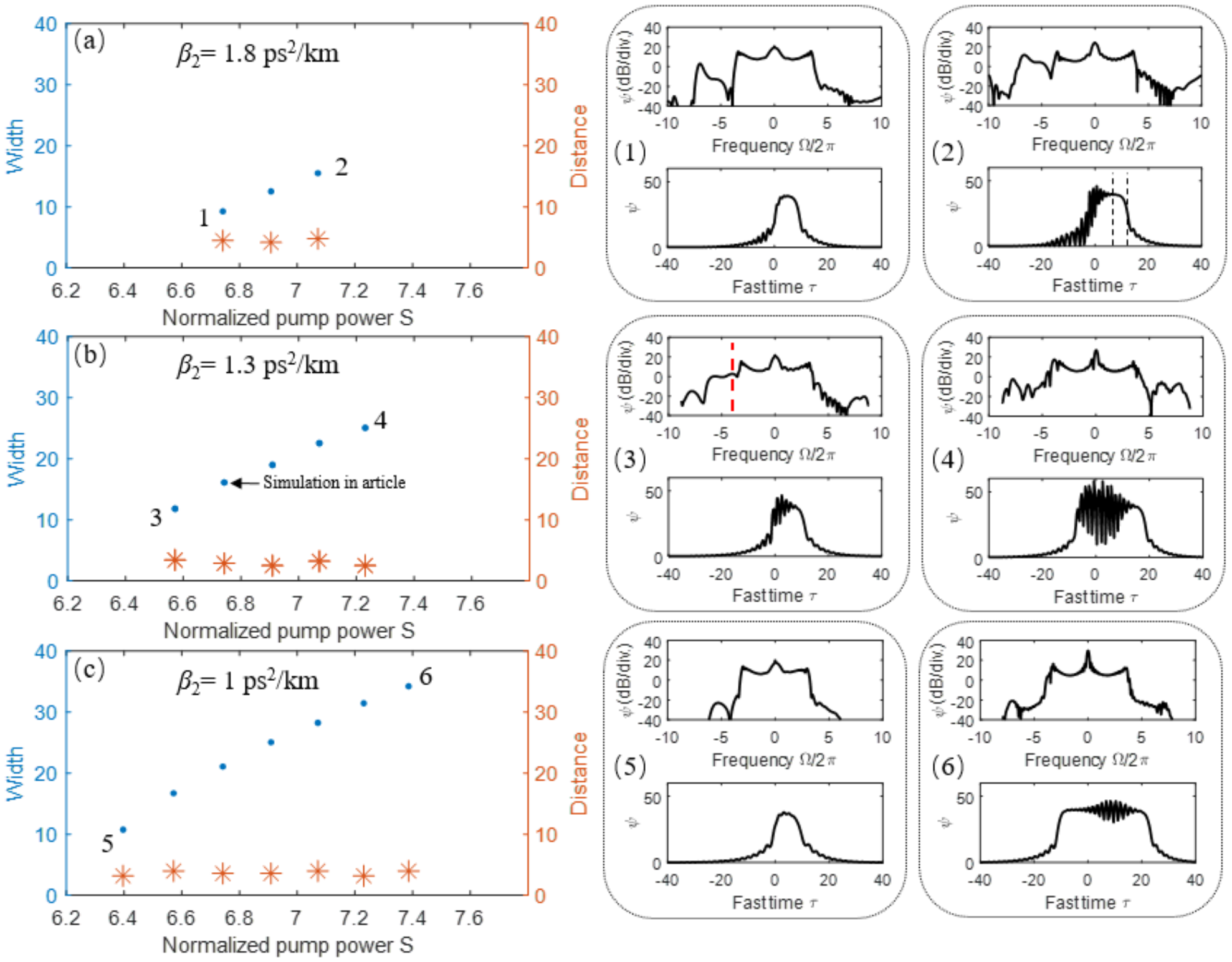

Fig. S2 Influence of GVD on the existence region of KRSs. (a)–(c) Normalized pulse width (dots) of the KRSs and the temporal distance (stars) between the top oscillation and down-SW with respect to the normalized pump power $S$ and different GVD values. (1)–(6) The corresponding spectra and temporal waveforms denoted in (a)-(c). The position of Raman gain spectrum is marked by the red dashed line in (3).

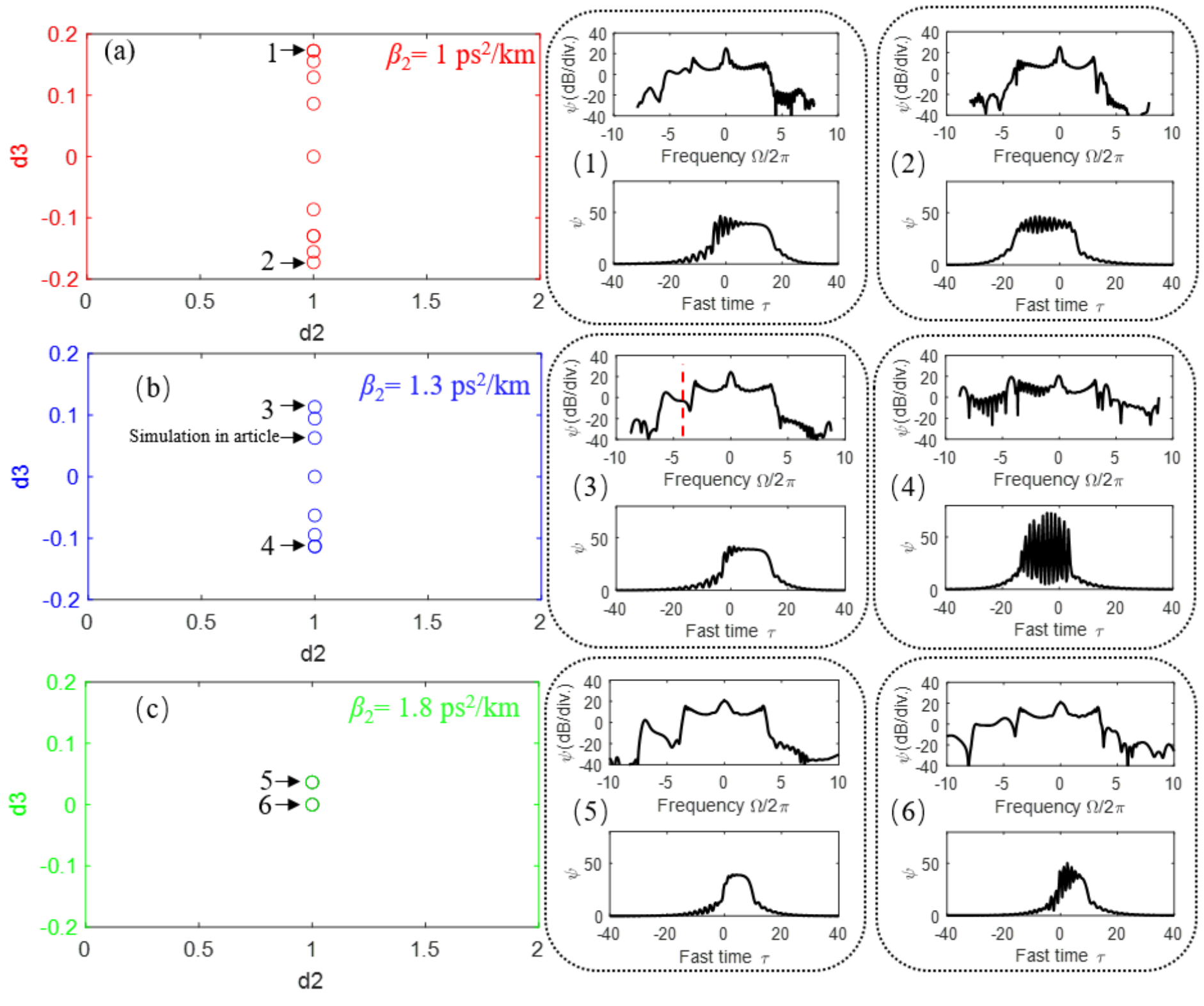

Fig. S3 Regions of TOD supporting KRSs under different GVD values of (a) 1 ps$^2$/km, (b) 1.3 ps$^2$/km and (c) 1.8 ps$^2$/km. (1)-(6) The corresponding spectra and temporal waveforms denoted in (a)-(c). The position of Raman gain spectrum is marked by the red dashed line.

## S3 Top oscillation locked to the down-SW in KRS states

In Fig. S2(a-c), normalized pulse width (dots) of the KRSs and the temporal distance (stars) between the top oscillation and down-SW with respect to different pump power and GVD are plotted. It can be clearly observed that the distance between the top oscillation and the down-SW (defined as the distance between the right edge of the top oscillation and the middle of the pulse falling edge, see dashed lines in Fig. S2(2)) is nearly constant with increasing power S for a given GVD. This indicates that this SRS effect induced top oscillation is locked to the down-SW which is consistent with theoretical prediction [2] and is a unique feature of KRS. In comparison, TOD-enabled zero-dispersion solitons also have top oscillation but it is locked to the up-SW [6].

## S4 Measurement of native rate (FSR) of the FP resonator

For pulsed pumping, it is crucial to measure the native rate (i.e., FSR) of the FP resonator. In this section, we present a dual-cavity structure which differs from the sweeping laser method limited by laser wavelength resolution. The setup is shown in Fig. S4(a). The FP resonator is placed in a fiber ring cavity containing an erbium-doped fiber amplifier (EDFA), an isolator (ISO), a polarization controller (PC), a 20-km-long single-mode fiber (SMF), and a 1% coupler. A 50-GHz photodetector (PD) is used to measure the beat note of the signal. Each mode of the FP cavity (linewidth 19 MHz) is discretely optical frequency sampled by the FSR (10 kHz) of the outer cavity. The harmonic beat note (Fig. S4(b)) is measured to be 3.678687 GHz, so the native rate of cavity is determined as 3.678687 / 2= 1.8393435 GHz. The accuracy of the device is not determined by the FSR of the outer fiber cavity, but rather by the frequency separation of the two orthogonal modes due to fiber birefringence, which is measured to be ~1 kHz [7].

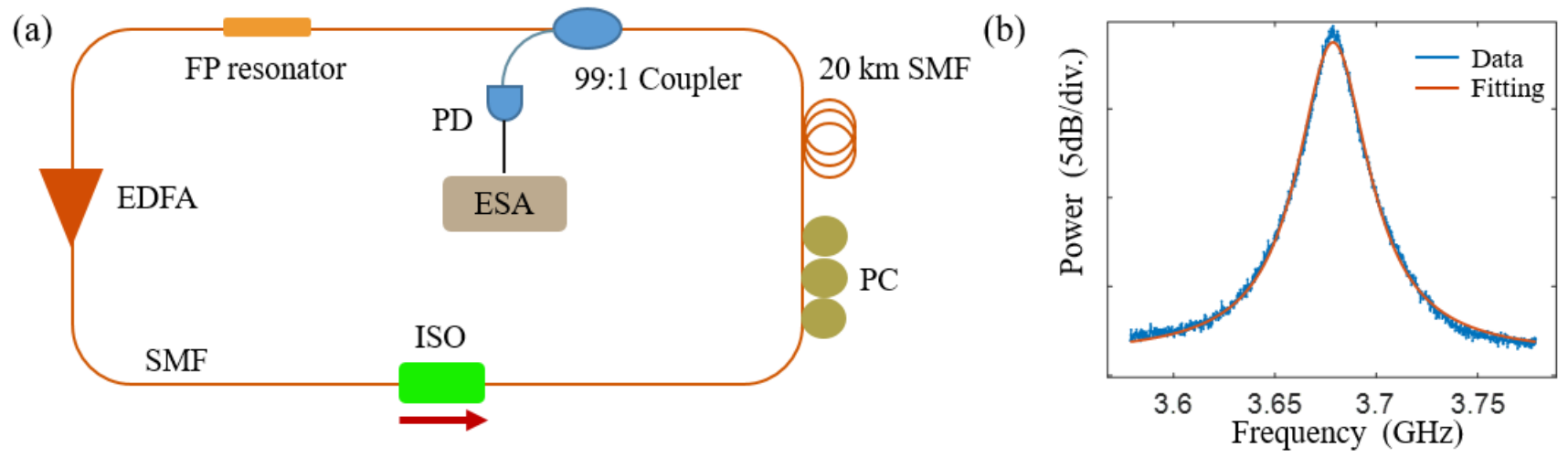

Fig. S4 Measurement of the native rate of FP resonator. (a) Experimental setup. PD: photodetector; ESA: electrical spectrum analyzer; PC: Polarization controller; ISO: isolator; EDFA: erbium-doped fiber amplifier. (b) Beat note at 2FSR of the FP resonator.

## S5 Dispersion measurement of the FP resonator

The dispersion measurement of the FP resonator is based on a fiber Mach-Zehnder interferometer (MZI) setup, as shown in Fig. S5(a) [8]. The reflection of the FP resonator, from which the resonance modes are recorded, is measured using a sweeping laser (Santec TSL-710) scanning from 1500 nm to 1560 nm at a rate of 80 nm/s. The resonance modes of the transmission light passing through the MZI are recorded simultaneously. The fiber length difference between the two unbalanced arms of the MZI is 20 m, corresponding to a frequency period $\Delta f_{MZI}$ ($\Delta f_{MZI}$= c/ $n_{eff}$ ΔL) of ~10 MHz. The $\Delta f_{MZI}$ at different wavelengths is also affected by the higher order dispersion, so we calibrated the laser swept wavelength and the higher order dispersion of MZI using two standard gas reference cells (HCN-13-50, $C_2H_2$ Wavelength References). The resonance modes $\omega_\mu$ of the FP resonator in the scanned wavelength range are calculated by the calibrated MZI. The dispersion profile $D_{int}(\mu)$ ($D_{int}(\mu) = \omega_\mu - \omega_0 - D1 \approx \mu^2 D_2/2 + \mu^3 D^3 /6$) is plotted in Fig. S5(b), with the GVD and TOD of the FP resonator fitted to be -28.2 Hz and 5.2 mHz, respectively. All devices used are placed in a constant temperature (uncertainty 0.01℃) chamber to avoid the temperature perturbation of the MZI and to meet the temperature requirements of gas cells.

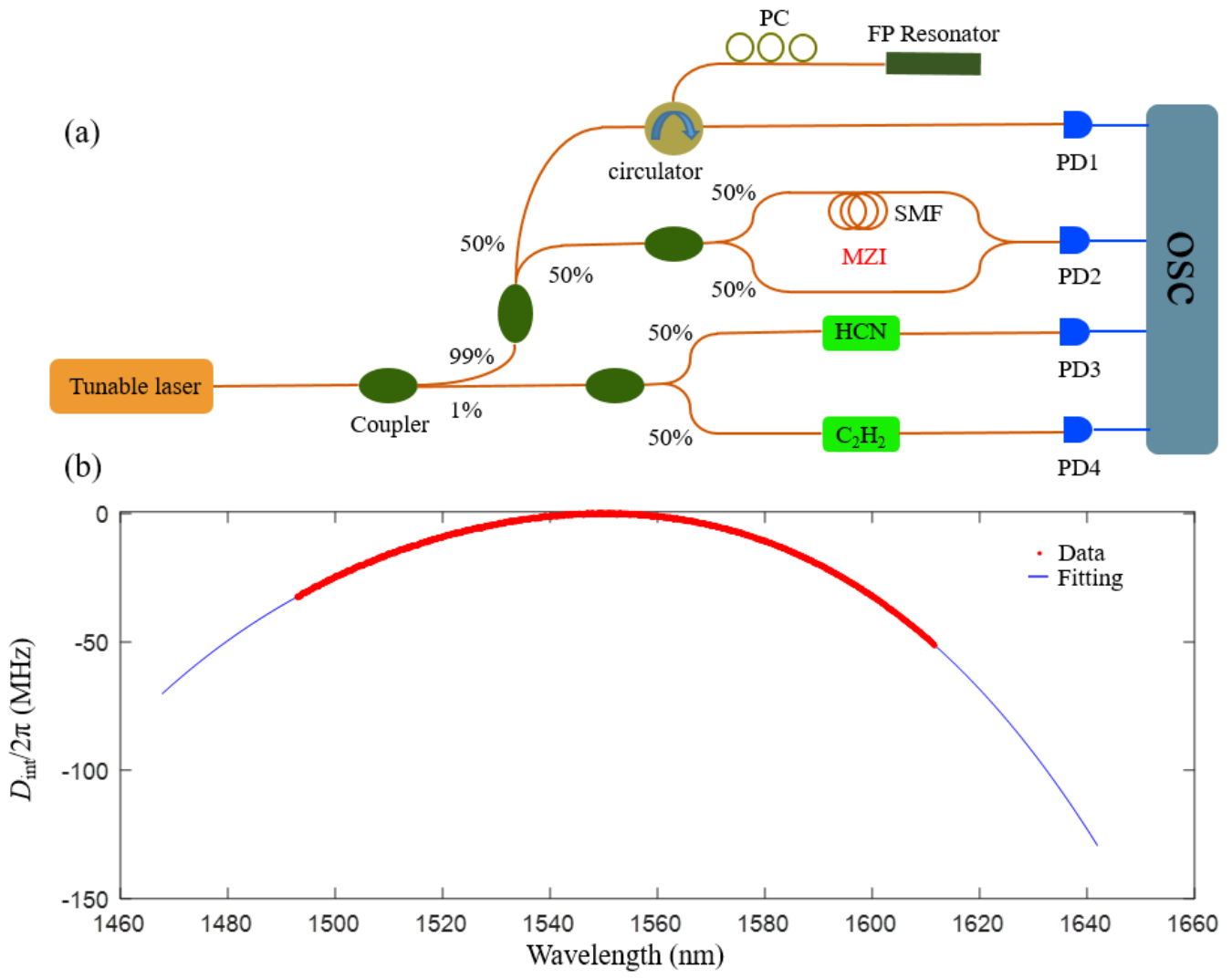

Fig. S5 Dispersion measurement of the FP resonator. (a) Experimental setup. SMF: single mode fiber; OSC: oscilloscope; PD: photodetector. (b) Measured dispersion curve of $D_{int}$

## S6 Performance comparison between KRS comb and other platicon works

The performance of our work is compared with other platicon works in normal GVD microcavities, as shown in Table S1. It can be observed that the generated KRS frequency comb of our work has several advantages: First, the large bandwidth (several times more than other works) can meet the needs of different wavelength bands. Second, the repetition rate of 1-10 Gigahertz can reduce PD detection rate requirements while the large number of comb lines greatly improve the resolution of detection for dual-comb spectroscopy or other applications [9]. Also, the native rate of the cavity allows a pulsed pump method to stimulate normal GVD optical frequency combs which avoids complex mode design (e.g., avoided mode crossing) or structural design (e.g., coupled resonators). Third, our platform is based on a fiber FP resonator, which has the advantages of simple coupling and easiness to achieve flat and low GVD over a wide wavelength range.

Table S1 Performance comparison with other experimental works in normal GVD microcavities

| Resonator type | Stimulation method | $\beta_2$ ($D_2$) | Repetition frequency | Bandwidth (-40 dB from pump) | Comb line numbers |
|---|---|---|---|---|---|
| $Si_3N_4$ microring [8] | Intensity modulated pump | 55 ps$^2$/km -180 kHz | 19.549 GHz | 1565-1615 (50 nm) | 320 |
| $Si_3N_4$ microring [10] | Mode interactions | 12.3 ps$^2$/km (-225 kHz) | 115.56 GHz | 1525-1675 (150 nm) | 163 |
| $Si_3N_4$ microresonator [11] | Photonic molecules | 39.7 ps$^2$/km (-600 kHz) | 104.84 GHz | 1500-1580 (80 nm) | 96 |
| DFB laser+$Si_3N_4$ microring [12] | self-injection locking | 63.5 ps$^2$/km (−59.9 kHz) | 26.2 GHz | 1545-1565 (20 nm) | 96 |
| Fiber ring resonator [13] | Pulsed pump | 7.5 ps$^2$/km (−17.7 Hz) | Integer multiple of 0.54 GHz | 1535-1565 (30 nm) | 6945 @ 0.54 GHz |
| **Fiber FP resonator (this work)** | **Pulsed pump** | **1.0 ps$^2$/km (-28.2 Hz)** | **3.6787 GHz** | **1440-1700 (260 nm)** | **9027** |